\documentclass[12pt]{iopart}

\usepackage{citesort}
\usepackage{pdfpages}
\begin{document}

\title[Intense narrowband XUV pulses from a compact setup]{Intense narrowband XUV pulses from a compact setup}

\author{M. Kretschmar, M. J. J. Vrakking and B. Schütte}
\address{Max-Born-Institut, Max-Born-Strasse 2A, 12489 Berlin, Germany}
\ead{Bernd.Schuette@mbi-berlin.de}
\vspace{10pt}
\begin{indented}
\item[]June 2021
\end{indented}

\begin{abstract}
We report on a compact and spectrally intense extreme-ultraviolet (XUV) source, which is based on high-harmonic generation (HHG) driven by 395~nm pulses. In order to minimize the XUV virtual source size and to maximize the XUV flux, HHG is performed several Rayleigh lengths away from the driving laser focal plane in a high-density gas jet. As a result, a high focused XUV intensity of $5 \times 10^{13}$~W/cm$^2$ is achieved, using a beamline with a length of only two meters and a modest driving laser pulse energy of 3~mJ. The high XUV intensity is demonstrated by performing a nonlinear ionization experiment in argon, using an XUV spectrum that is dominated by a single harmonic at 22~eV. Ion charge states up to Ar$^{3+}$ are observed, which requires the absorption of at least four XUV photons. The high XUV intensity and the narrow bandwidth are ideally suited for a variety of applications including photoelectron spectroscopy, the coherent control of resonant transitions and the imaging of nanoscale structures.
\end{abstract}
\maketitle


Ultrashort, intense extreme-ultraviolet (XUV) pulses are used in a growing number of applications, which include nonlinear multiphoton ionization of atoms~\cite{wabnitz05,sorokin07,motomura09,nayak18,bergues18,senfftleben20}, molecules~\cite{sorokin06,jiang09} and clusters~\cite{wabnitz02,iwayama10,schutte14a}, second-harmonic generation in thin films~\cite{helk21} and the study of XUV strong-field physics~\cite{ott19,ding19}. Intense XUV pulses are also a prerequisite for performing XUV-pump XUV-probe experiments where XUV pulse durations down to the attosecond regime have already been used~\cite{tzallas11,takahashi13,schnorr14}. While the study of electron dynamics on these extremely short timescales requires broadband XUV pulses, intense XUV pulses with a narrower bandwidth are advantageous for a range of applications including photoelectron spectroscopy~\cite{meyer10,varvarezos21} and the study of resonant transitions, e.g. within four-wave mixing~\cite{bencivenga15}, superfluorescence~\cite{harries18} and the control of Rabi oscillations~\cite{flogel17}. Furthermore, XUV pulses with a narrow bandwidth and a high spectral intensity are ideally suited for single-shot coherent diffractive imaging (CDI) of nanostructures and nanoscale targets~\cite{ravasio09,bostedt12,rupp17}. 

Narrowband intense XUV pulses are available from free-electron laser (FEL) facilities~\cite{ackermann07,shintake08,allaria12}, but the limited access and the large size of these facilities can make experiments very challenging. Alternatively, long high-harmonic generation (HHG) beamlines with lengths around 10 meters or even longer have been developed for the generation of intense XUV pulses~\cite{mashiko04,tzallas11,takahashi13,schutte14a,manschwetus16,bergues18,nayak18,senfftleben20}. Intense XUV sources based on HHG are currently being developed, including the user facilities ELI beamlines in Prague~\cite{hort19} and ELI-ALPS in Szeged~\cite{kuhn17}. A disadvantage of these sources is that they require very powerful laser systems for driving the HHG process, often reaching the multi-terawatt range. At the same time, these large-scale setups lead to high demands regarding the laser stability. This can be challenging, especially when considering that the XUV pulses are typically focused to micrometer or even nanometer spot sizes~\cite{motoyama19,major20} and that some of these experiments require attosecond stability. A number of more compact XUV sources have been reported that were used to study two-photon ionization and absorption in He, resulting in the generation of singly-charged He ions~\cite{kobayashi98,sekikawa04,barillot17}.

In state-of-the-art setups devoted to the generation of intense XUV pulses based on HHG, the focus often lies on maximizing the XUV flux by using powerful driving lasers and by loosely focusing the driving laser pulses into the HHG medium~\cite{kuhn17,nayak18,bergues18,li20}. Recently, we have used a different approach, demonstrating that optimization of the XUV intensity on target requires a choice of parameters entirely different from the parameters needed to optimize the XUV pulse energy~\cite{senfftleben20}. This approach was based on using a modest focal distance ($\geq5$~m) for the near-infrared (NIR) driving laser, followed by a long propagation distance of the generated XUV beam. This enables large demagnification of the XUV source size and resulted in a high XUV intensity of $7 \times 10^{14}$~W/cm$^2$. Using these pulses for multiphoton ionization, charge states up to Ar$^{5+}$ were observed following the absorption of at least 10 XUV photons. At the same time, a moderate NIR pulse energy of 11~mJ was used~\cite{senfftleben20}. However, this setup still required a lot of space, since overall an 18-m-long beamline was used.

An important consideration for the generation of intense XUV pulses is that optical elements should only be used where absolutely necessary, because these typically suffer from high reflection / transmission losses and / or aberrations. Furthermore, only modest demagnification of the XUV source size can be achieved even in complex optical arrangements, e.g. when three toroidal mirrors are used~\cite{poletto13}. Instead, a promising path is to exploit the inherent properties of the generated XUV pulses. Recently, we demonstrated that generating high harmonics several Rayleigh lengths away from the driving laser focus can result in the generation of a large high-harmonic generation volume and a small virtual XUV source size of only a few micrometers. After refocusing the XUV pulses, a high XUV intensity of $2 \times 10^{14}$~W/cm$^2$ was achieved in a compact setup with a length of only 2 meters~\cite{major21}, enabling triple ionization of argon atoms. In these experiments the XUV pulses used for the ionization of argon atoms consisted of four different harmonic orders, which made it difficult to understand the ionization pathways in detail. 

Here we demonstrate the generation of intense narrowband XUV pulses in a simple and compact setup, where high harmonics are generated several Rayleigh lengths away from the driving laser focus in a short, high-density gas jet. The HHG is driven by the second harmonic of the fundamental NIR pulses, resulting in a large spectral separation between the individual harmonic orders. Using an Al filter to block the driving laser and a normal-incidence XUV focusing mirror, an XUV pulse dominated by a single harmonic order at 22~eV is obtained. In the ionization of Ar atoms, Ar$^{2+}$ and Ar$^{3+}$ ions are observed, which requires the absorption of at least two and four XUV photons, respectively.


\begin{figure}
 \centering
  \includegraphics[width=8.6cm]{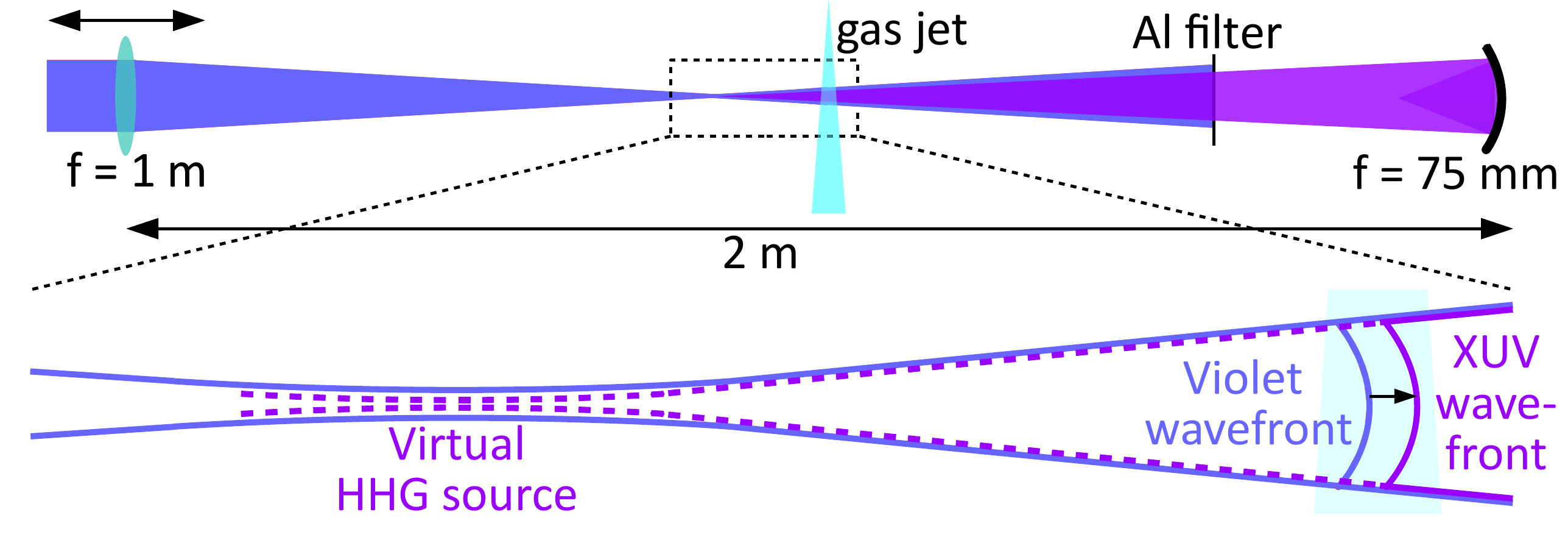}
 \caption{\label{figure_setup} Compact intense XUV setup. A violet laser pulse at 395~nm is focused by a lens with a focal length of 1~m. High harmonics are generated in a gas jet that is placed several Rayleigh lengths away from the focal plane of the violet laser. As a result, both the size and the divergence of the generated XUV beam are similar to the size and the divergence of the driving laser. However, as shown in the zoomed region in the lower part of the graph, the virtual XUV source size is significantly smaller than the focal size of the violet laser, as a consequence of the shorter wavelength of the XUV beam. An Al filter is used to attenuate the violet laser beam. High XUV intensities are obtained by demagnifying the XUV source size using an XUV focusing element with a short focal length $f$. To this end, a spherical B$_4$C-coated mirror with $f=75$~mm is used. }
\end{figure}

The concept for generating intense XUV pulses in a compact setup is presented in Fig.~1. First, NIR pulses centered at 790~nm with an energy of 13~mJ and a duration of 40~fs~\cite{gademann11} were frequency-doubled in a 150-$\mu$m-thick beta-barium borate (BBO) crystal by second-harmonic generation (SHG), resulting in a pulse energy of 3~mJ at 395~nm. Two dichroic mirrors were used to separate the generated violet beam from the fundamental beam. Next, as shown in Fig.~1, these pulses were focused using a spherical lens with a focal length of 1~m, resulting in an estimated peak intensity of $8 \times 10^{15}$~W/cm$^2$ in the focal plane when no gas was present. High harmonics were generated in a high-density gas jet produced by a cantilever piezoelectric valve with a nozzle diameter of 0.5~mm~\cite{irimia09} that was placed either before or behind the focal plane of the violet laser. By generating high harmonics several Rayleigh lengths away from the focal plane of the violet laser, curved wavefronts are transferred from the driving laser to the generated XUV beam (see lower graph in Fig.~1). As a consequence, the violet and the XUV beams are expected to have similar divergences. Due to the shorter wavelength, however, the XUV virtual source size is significantly smaller than the focus size of the driving laser (see also Ref.~\cite{major21}). By further demagnifying the virtual XUV source size using a spherical mirror (coated with B$_4$C) with a short focal length of 75~mm, a high XUV intensity can be achieved using a setup that has a length of only 2 meters. An important advantage of generating harmonics in this scheme is the large generation volume leading to a comparably high XUV photon flux. 

A 100-nm-thick Al filter was used to block the driving laser, while transmitting XUV photons with energies higher than about 16~eV. The XUV beam profile was recorded using a microchannel plate (MCP) / phosphor screen assembly that was placed at a distance of about 50~cm from the gas jet. For the nonlinear ionization of argon, a pulsed atomic jet of argon atoms was generated by a second cantilever piezoelectric valve with a nozzle diameter of 0.5~mm. The central part of the atomic jet was selected by a skimmer with an orifice diameter of 0.5~mm. Following interaction of the focused XUV beam with the atomic beam, the generated ions were recorded using a velocity-map imaging spectrometer (VMIS)~\cite{eppink97} that was operated in spatial-map imaging mode~\cite{stei13}. Individual ion traces were separately recorded by gating the detector and changing the delay of the gating window.  


\begin{figure*}[tb]
 \centering
  \includegraphics[width=12.9cm]{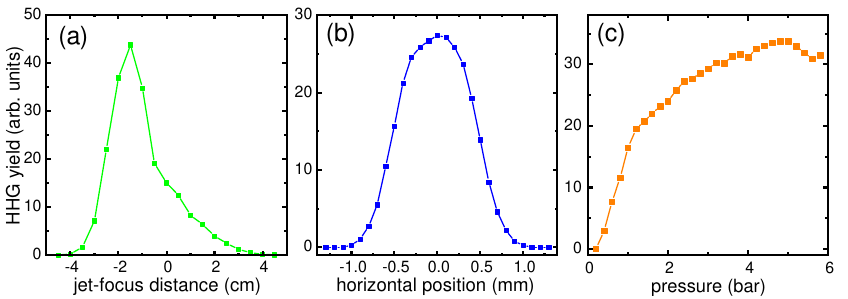}
 \caption{\label{HHGproperties} (a) HHG yield obtained in Kr using a backing pressure of 4~bar as a function of the distance between the gas jet and the focal plane of the driving laser. Negative distances correspond to the gas jet being placed behind the driving laser focus. The laser intensity of the violet beam at the focus when the gas jet is switched off is estimated as $8 \times 10^{15}$~W/cm$^2$. (b) HHG yield as a function of the transverse position of the gas jet at a jet-focus distance of $-1.5$~cm and a backing pressure of 4~bar. (c) HHG yield as a function of the backing pressure at a jet-focus distance of $-1.5$~cm.}
\end{figure*}

Fig.~2(a) shows the dependence of the obtained HHG yield in Kr using a backing pressure of 4~bar on the distance between the focal plane of the driving laser and the gas jet. This curve has a maximum at $-1.5$~cm, which means that the focus is located approximately 5 Rayleigh lengths before the gas jet. Under these conditions, both the XUV beam size at the jet and its divergence are expected to be similar to the size and the divergence of the driving laser (see Fig.~1 and Ref.~\cite{major21}). To obtain information about the width of the gas jet, the dependence of the HHG yield on the transverse position of the gas jet with respect to the driving laser beam was recorded, as presented in Fig.~2(b). This curve has a full width at half maximum (FWHM) of about 1~mm. Fig.~2(c) depicts the relative HHG yield as a function of the backing pressure, which exhibits a steep increase up to about 1~bar. Saturation starts to set in at higher pressures, and from about 4~bar the HHG yield is almost constant. This is comparable to our previous results using 800~nm driving pulses, where this behavior was attributed to reshaping of the fundamental laser and its influence on phase-matching under these conditions~\cite{major21}. The saturation behavior is advantageous, since it makes the HHG optimization comparably simple and because slightly different pressures experienced by the driving laser across its beam profile have a negligible effect on the spatially dependent HHG yield. From Fig.~2(c) one can extract that the HHG yield at 2~bar is about 77~$\%$ of the HHG yield at 4~bar. Using this information in combination with the results shown in Fig.~2(b), we estimate that the region of the gas jet where the yield drops to 77~$\%$ on both sides of the maximum (meaning that the pressure has dropped by a factor of about 2) spans a width of about 0.75~mm. This represents an estimation of the gas jet width and is in accordance with expectations, considering that the nozzle diameter of the gas valve is 0.5~mm and that the exit of the nozzle is placed as close to the laser beam as possible.   

\begin{figure}[htb]
 \centering
  \includegraphics[width=7cm]{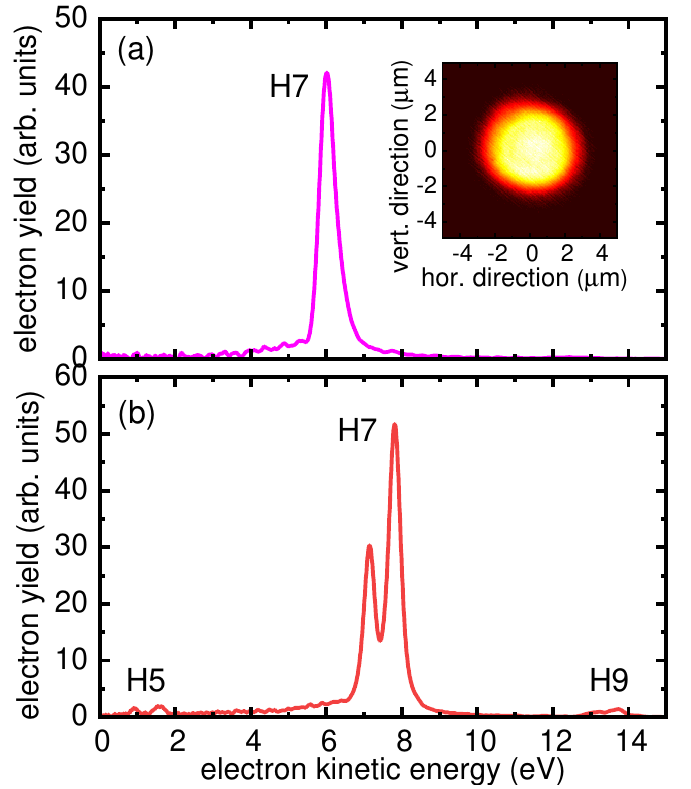}
 \caption{\label{figure_beam-spectrum} Photoelectron spectra measured for the ionization of (a) Ar and (b) Kr after focusing by the B$_4$C-coated spherical mirror. The spectra are dominated by contributions from the 7th harmonic (H7) at a photon energy of 22~eV. Spin-orbit splitting is visible in the photoelectron spectrum of Kr. The inset in (a) shows the XUV beam profile recorded 50~cm behind the gas jet. }
\end{figure}

The XUV beam profile recorded at a distance of 50~cm behind the driving laser focus is presented as an inset in Fig.~3(a). The beam radius at this position is about 3~mm, corresponding to an FWHM divergence of 7~mrad, which is similar to the divergence of the violet laser and is a direct consequence of performing the HHG process several Rayleigh lengths away from the focal plane of the driving laser. In Fig.~3(a), a photoelectron spectrum is shown that was obtained after focusing the XUV beam into an Ar gas jet. The spectrum is dominated by the contribution from a single harmonic order with a photon energy of about 22~eV, corresponding to the 7th harmonic of the violet driving laser (i.e. the 14th harmonic of the NIR laser). For comparison, Fig.~3(b) shows a photoelectron spectrum obtained for the ionization of Kr, showing that the contributions from both the 5th and the 9th harmonics are small. The dominance of a single harmonic order was facilitated by the large spectral separation between the harmonic orders ($\approx 6.3$~eV) and by exploiting the filtering characteristics of the Al filter (which has a high transmission for photon energies $>16$~eV~\cite{henke93}) and the broadband boron carbide focusing mirror (which has a high reflectivity for photon energies $<25$~eV~\cite{larruquert99}). The XUV pulse energy at the source was approximately 350~nJ, as measured by an XUV photodiode (AXUV100G, Opto Diode). Taking into account the transmission through the Al filter (40~$\%$) and the reflectivity of the XUV focusing mirror (25~$\%$), an XUV pulse energy of 35~nJ is estimated on target. The XUV pulse duration is estimated as 25~fs, considering that the nonlinear HHG process typically leads to an XUV pulse duration which is approximately half the driving laser pulses duration, see e.g.~\cite{mauritsson04,nabekawa05}. An XUV pulse duration of 25~fs was also obtained in HHG simulations when using 800~nm driving pulses~\cite{major21}.


\begin{figure}[htb]
 \centering
  \includegraphics[width=7cm]{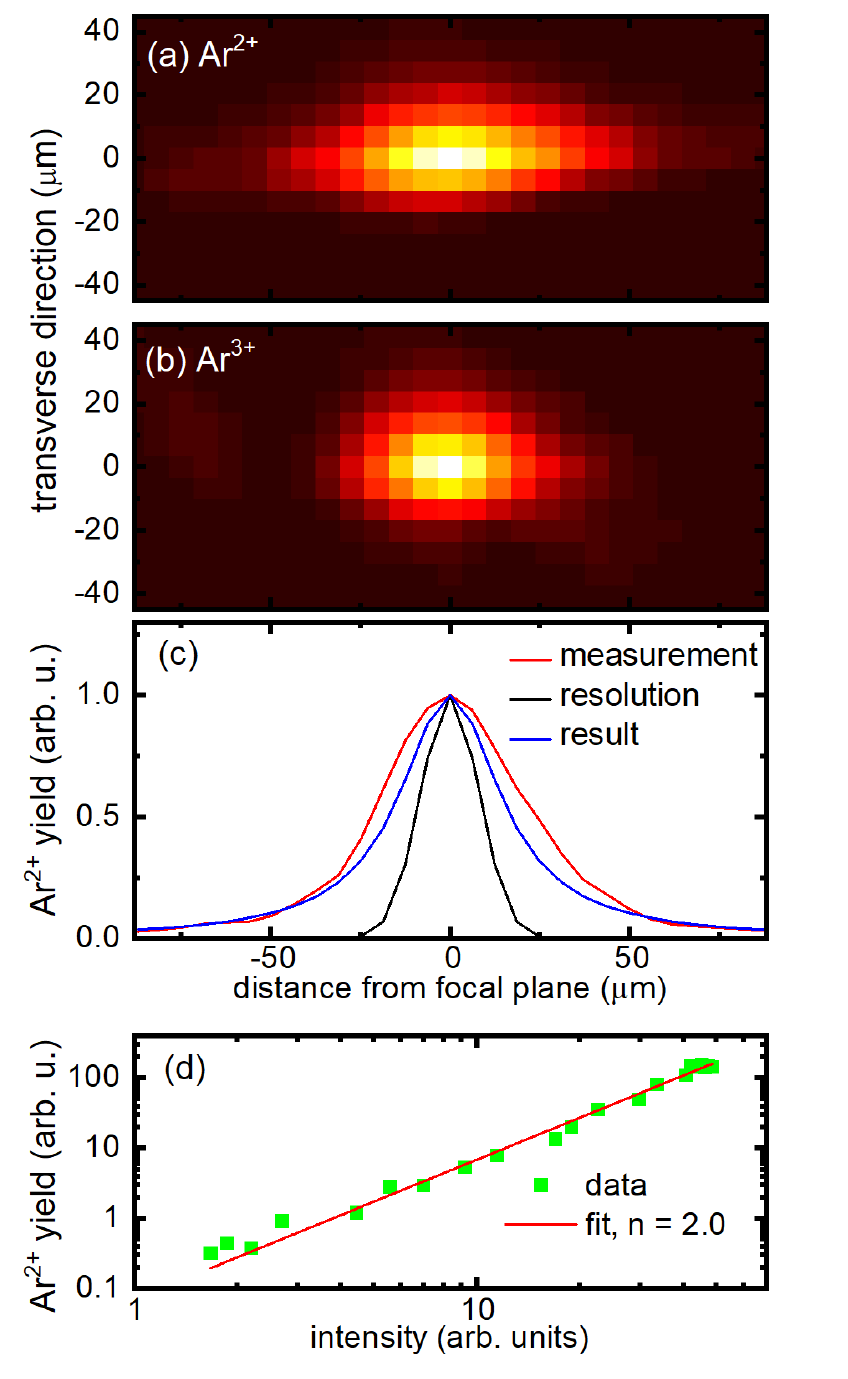}
 \caption{\label{figure_Ar2+} Distributions of (a) Ar$^{2+}$ and (b) Ar$^{3+}$ ions as a function of the distance from the XUV focal plane. The Ar$^{3+}$ ion distribution is narrower because of the higher degree of nonlinearity involved. (c) Integrated Ar$^{2+}$ ion yield (red curve), spatial resolution (black curve) and their deconvolution (blue curve). See text for details. (d) Ar$^{2+}$ ion yield $Y$ as a function of the XUV intensity (green squares) and a fit according to $Y\propto I^n$ (red curve), where $I$ refers to the XUV peak intensity. A nonlinearity of $n=2.0$ is obtained.}
\end{figure}

To demonstrate the high intensity of our XUV source, we performed a nonlinear ionization experiment in Ar atoms. This target was chosen, because it allows a comparison with other experiments performed at FELs~\cite{wabnitz05,motomura09} and using intense HHG sources~\cite{nayak18,senfftleben20}. To this end, the XUV pulses were focused using a spherical mirror with a focal length of $f=75$~mm. As shown in Fig.~4(a)+(b), both Ar$^{2+}$ and Ar$^{3+}$ ions were generated. The energy required to generate Ar$^{2+}$ is 43.4~eV, and the energy required to generate Ar$^{3+}$ is 84.1~eV~\cite{NIST}, meaning that at least two and four XUV photons need to be absorbed to generate Ar$^{2+}$ and Ar$^{3+}$, respectively. In Fig.~4(a)+(b), the spatial distributions of the different ion species are presented as a function of the distance from the XUV focal plane. These distributions are centered at the XUV focal plane, reflecting the nonlinear nature of the ionization process. The narrower distribution of the Ar$^{3+}$ ion trace compared to the Ar$^{2+}$ ion trace is a consequence of the higher order of the nonlinearity involved in the former case. The spatially resolved ion traces allow us to determine the relative ion yields at the XUV focal plane, giving an Ar$^{2+}$ / Ar$^+$ ion yield ratio of 8~$\%$ and an Ar$^{3+}$ / Ar$^{2+}$ ion yield ratio of 17~$\%$.

The transversely integrated Ar$^{2+}$ ion yield shown in Fig.~4(c) (red curve) is used to estimate the XUV Rayleigh length $z_R$. According to Fig.~4(d), where the Ar$^{2+}$ yield is shown as a function of the XUV intensity, the generation of Ar$^{2+}$ is a two-photon process. Therefore, the Ar$^{2+}$ ion yield as a function of the distance $z$ from the XUV focal plane is proportional to $(1+z^2/z_R^2)^{-1}$~\cite{major21}. The measured ion distribution is further influenced by the spatial resolution of our method, see black curve in Fig.~4(c). The spatial resolution was estimated by the transverse width of the Ar$^{2+}$ ion distribution at the XUV focal plane. After performing a deconvolution, the blue curve is obtained, which yields a Rayleigh length of 17~$\mu$m. This allows us to estimate the XUV beam waist radius according to $w_0 = w_{XUV,mirror} (1 + d_{XUV}^2/z_R^2)^{-1/2} =(1.3\pm 0.2)$~$\mu$m, where $w_{XUV,mirror}=6$~mm is the estimated XUV beam radius on the focusing mirror and $d_{XUV}=81$\,mm is the distance between the mirror and the imaging plane. We note that an imperfect alignment of the spherical mirror leading to astigmatism might affect the estimate of $w_0$. Using $w_0=1.3$~$\mu$m, an XUV peak intensity of $5 \times 10^{13}$~W/cm$^2$ would be obtained. Our results are thus consistent with previous FEL results reported at a photon energy of 20~eV, where Ar$^{3+}$ was the highest charge state at an intensity of $5 \times 10^{13}$~W/cm$^2$~\cite{motomura09}. In our previous experiment performed at an 18-m-long HHG beamline, Ar$^{3+}$ ions were only observed at intensities of $1 \times 10^{14}$~W/cm$^2$ and higher~\cite{senfftleben20}. The higher intensity required for the observation of Ar$^{3+}$ in that experiment can be attributed to the much shorter XUV pulse duration of about 3~fs in that case~\cite{major20}, meaning that significantly less XUV photons are absorbed before the peak of the XUV intensity is reached. In contrast to the previously mentioned results, Ar$^{3+}$ was only observed at XUV intensities exceeding $1 \times 10^{15}$~W/cm$^2$ using the intense XUV source based on HHG reported in Ref.~\cite{nayak18}. This result was attributed to the higher degree of coherence of HHG sources compared to FEL sources, but is not corroborated by the results reported here.

In summary, we have reported on the generation of intense narrowband XUV pulses in a compact and simple setup, which is driven by 395~nm pulses with a moderate pulse energy of 3~mJ. It is straightforward to implement this concept in many laboratories, since no large laboratories or multi-terawatt laser systems are needed, in contrast to many state-of-the-art intense XUV sources based on HHG~\cite{manschwetus16,takahashi13,nayak18,bergues18,senfftleben20,hort19,kuhn17,li20}. Our results could therefore boost research fields that require intense XUV pulses such as XUV strong-field physics and coherent diffractive imaging of nanoscale particles and nanostructures, thereby overcoming the challenges experienced in approaches where several harmonic orders are used~\cite{rupp17}. In comparison to setups where a multilayer XUV mirror is used to select a single harmonic order~\cite{ravasio09}, an advantage of the scheme presented here is the higher pulse energy that is contained in each harmonic order. By spectrally tuning either the fundamental laser or the second harmonic, tunable intense XUV pulses can be obtained and can be used to scan across resonances, similar to the way that this can be done at FELs~\cite{takanashi17}. Instead of the second harmonic, it is also possible to use the third harmonic for HHG~\cite{popmintchev15}, allowing access to different harmonic orders and resulting in an even larger spacing between the individual harmonic orders.

\section*{References}
\bibliography{Bibliography}

\end{document}